**Title**: Fully Automated Deep Learning-enabled Detection for Hepatic Steatosis on Computed Tomography: A Multicenter International Validation Study

**Authors**: Zhongyi Zhang[1], Guixia Li[2], Ziqiang Wang[3], Feng Xia[4], Ning Zhao[5], Huibin Nie[6], Zezhong Ye[7], Joshua Lin[8], Yiyi Hui[9]* & Xiangchun Liu[1]*

**Affiliations**:

1. Department of Nephrology, The Second Hospital of Shandong University, Shandong University, Jinan, 250033, Shandong, China
2. Department of Nephrology, Shenzhen Third People's Hospital, the Second Affiliated Hospital of Southern University of Science and Technology, Shenzhen, 518112, Guangdong, China
3. Department of Nephrology, The First Affiliated Hospital of Hainan Medical University, Haikou, 570102, Hainan, China
4. Department of Cardiovascular Surgery, Wuhan Asia General Hospital, Wuhan, 430000, Hubei, China
5. The First Clinical Medical School, Shanxi Medical University, Shanxi, 030001, Taiyuan, China
6. Department of Nephrology, Chengdu First People's Hospital, Chengdu, 610095, Sichuan, China
7. Independent Researcher, Boston, 02115, MA, USA
8. Keck School of Medicine, University of Southern California, Los Angeles, 90033, CA, USA
9. Department of Medical Imaging, Shandong Provincial Hospital Affiliated to Shandong First Medical University, Jinan, 250021, Shandong, China

* Corresponding authors with the same contribution
Xiangchun Liu, MD, PhD.
Department of Nephrology, The Second Hospital of Shandong University, Shandong University, Jinan, 250033, Shandong, China
Email: liuxiangchun@sdu.edu.cn

Yiyi Hui, MD.
Department of Medical Imaging, Shandong Provincial Hospital Affiliated to Shandong First Medical University, Jinan, 250021, Shandong, China
Email: huiyiyi@sdfmu.edu.cn

**Abstract**

Despite high global prevalence of hepatic steatosis, no automated diagnostics demonstrated generalizability in detecting steatosis on multiple international datasets. Traditionally, hepatic steatosis detection relies on clinicians selecting the region of interest (ROI) on computed tomography (CT) to measure liver attenuation. ROI selection demands time and expertise, and therefore is not routinely performed in populations. To automate the process, we validated an existing artificial intelligence (AI) system for 3D liver segmentation and used it to purpose a novel method: AI-ROI, which could automatically select the ROI for attenuation measurements. The AI segmentation and AI-ROI method were evaluated on 1,014 non-contrast enhanced chest CT images from eight international datasets: LIDC-IDRI, NSCLC-Lung1, RIDER, VESSEL12, RICORD-1A, RICORD-1B, COVID-19-Italy, and COVID-19-China. The automatic segmentation achieved a dice coefficient of 0.957 ± 0.046 (mean ± standard deviation). Attenuations measured by AI-ROI showed no significant differences ($p = 0.545$) and a reduction of 71% time compared to expert measurements. The area under the curve (AUC) of the steatosis classification of AI-ROI is 0.921 (95% CI: 0.883 - 0.959). If performed as a routine screening method, our AI protocol could potentially allow early non-invasive, non-pharmacological preventative interventions for hepatic steatosis. 1,014 expert-annotated liver segmentations of patients with hepatic steatosis annotations can be downloaded here: https://drive.google.com/drive/folders/1-g_zJeAaZXYXGqL1OeF6pUjr6KB0igJX.

**Abbreviations:** 2D = Two-dimensional, 3D = Three-dimensional, AI = Artificial intelligence, AUC = Area under the curve, COVID-19 = Coronavirus Disease 2019, CI = Confidence interval, DSC = Dice coefficient, HU = Hounsfield unit, NIfTI = Neuroimaging Informatics Technology Initiative, NLST = National Lung Screening Trial, ROI = Region of interest, std = Standard deviation.

**Introduction**

Hepatic steatosis, also called fatty liver disease, occurs when intrahepatic fat is ≥5% of liver weight[1]. Associated with diabetes, hypertension, obesity, and severe COVID-19 pneumonia[2], patients with hepatic steatosis experience increased risk for cirrhosis, end-stage liver failure, and early mortality[3]. Non-alcoholic fatty liver disease, a subtype of hepatic steatosis, is recognized as the most common cause of chronic liver disease in the West[4]. Early detection of steatosis might mitigate fat-related liver dysfunction, preventing disease progression that ultimately leads to end-stage liver failures.

Liver biopsy, an invasive procedure, remains the clinical diagnostic standard for hepatic steatosis but carries morbidity risks[5]. Given the constraint of invasive studies, non-invasive imaging methods seem the next-best diagnostics that may popularize early detection of hepatic steatosis while minimizing morbidity to patients. Non-contrast enhanced computed tomography (CT) has been proven to detect hepatic steatosis reliably in a non-invasive manner[6,7]. A liver attenuation of <= 40 HU in a clinician-selected region of interest (ROI) would generally indicate moderate-to-severe hepatic steatosis[8,9]. However, ROI selection demands time and expertise, so this detection method is not routinely performed. As hepatic steatosis remains a public health concern but is still unexplored in millions of CT images, including scans for lung cancer[10,11] and COVID-19[12,13] diagnosis, automated detections are needed.

Deep learning, a form of artificial intelligence (AI), is the frontier of automated medical image analysis[14]. It can automatically segment CT images and detect abnormalities within seconds[15,16]. Some deep learning methods were developed in a 3D manner to automatically detect hepatic steatosis[17–20], but no study automated the process of ROI selections and statistical differences existed compared to expert measurements. Furthermore, current studies[17–20] lack proven generalizability across multiple international datasets. Most importantly, none of these studies[17–20] released curated datasets, which could have served as development or validation data to promote deep learning algorithms and hepatic research.

In this work, we hypothesized that one method could automatically mimic experts to select ROIs for hepatic steatosis detection on multicenter international CT images. Thus, we explored automatic methods for ROI selection based on a 3D deep learning system[20]. We aimed to evaluate the new method across various CT scanners and imaging centers in clinical settings worldwide. If proven to be as accurate as human experts, the method could be applied worldwide to monitor fat-related dysfunctions and therefore prevent the disease from progressing to end-stage liver failures.

**Materials and Methods**

An overview of the study design is provided in Fig. 1.

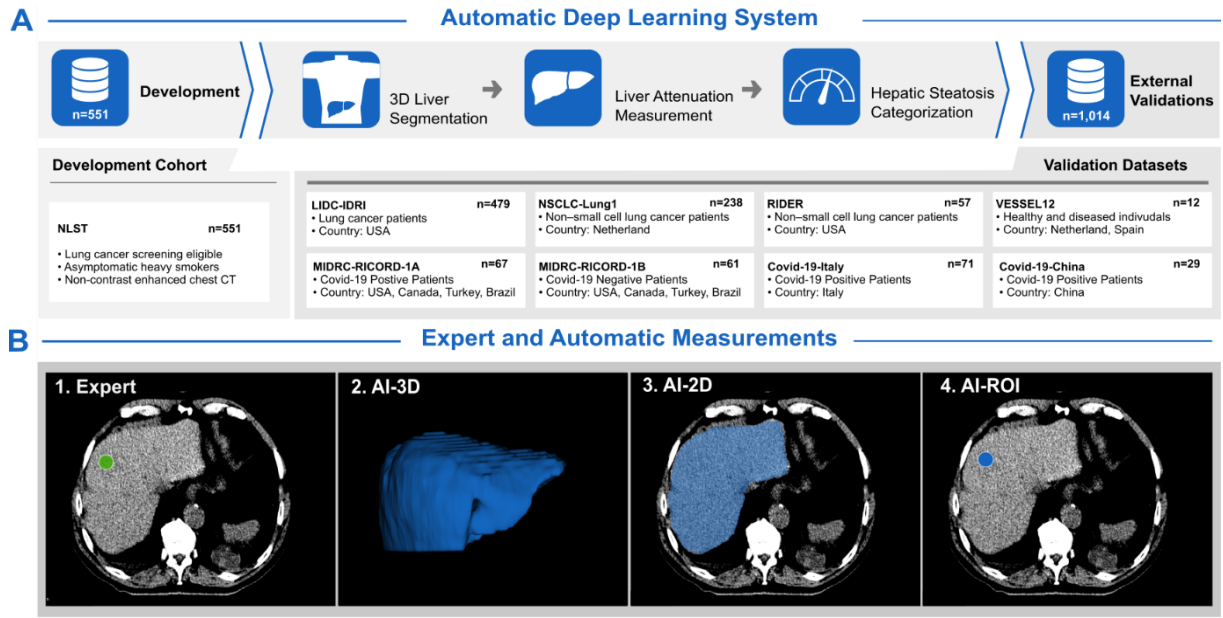

**Figure 1. A**: An existing deep learning system was validated on eight distinct datasets for hepatic steatosis classification in our study. The system was developed on chest CT images of asymptomatic heavy smokers from the NLST cohort in a previous study. **B**: Liver attenuation was measured by experts and artificial intelligence (AI) separately in Hounsfield unit (HU). 1. Expert: Experts measured the mean pixel value on the region of interest (ROI) of three cross-sectional images at different hepatic levels (green); 2. AI-3D: AI-measured mean voxel value on the automatic segmentation (blue) by the deep learning system directly in a 3D manner; 3. AI-2D: a novel post-processing method, AI-measured mean pixel value on a single axial slice containing the largest cross-sectional area in a 2D manner (blue); 4. AI-ROI: a novel post-processing method, AI-measured mean pixel value on the ROI (blue) of three cross-sectional images at different hepatic levels. Three slices refer to one slice with the largest cross-sectional area and two neighborhood slices 5 mm away from the center slice. One ROI selected by an expert or AI in one center slice was shown as an example. NLST = National Lung Screening Trial.

**Study Design and Datasets**

This retrospective study aims to purpose a fully automatic method to select the ROI on CT to predict the existence of hepatic steatosis (Fig. 1). This study was conducted in accordance with the Declaration of Helsinki guidelines and received approval from the local institutional review board. A waiver of consent was obtained from the institutional review board as the analyzed data came from public datasets or conducting a retrospective study.

For a multicenter international validation, 1,014 CT images were collected from 8 publicly available, de-identified patient datasets. Each dataset was downloaded and curated from the following databases: LIDC-IDRI[21], NSCLC-Lung1[22], RIDER[23], VESSEL12[24], MIDRC-RICORD-1A[25], MIDRC-RICORD-1B[25], COVID-19-Italy[26], and COVID-19-China[27]. Four datasets[25–27] were initially obtained to diagnose COVID-19, three datasets[21–23] were scanned for lung cancer detection, and one dataset[24] was released for lung vessel segmentation. Non-contrast enhanced chest CT image was the only inclusion criterion. The following image modalities were thus excluded: (i) contrast-enhanced chest CT images, which were selected out by an automatic method[28] and then validated by a radiologist; (ii) stomach with contrast agent; (iii) imaged liver with severe artifacts; livers with tumors or cysts collectively bigger than 5 cm$^3$. Fig. 2 summarizes the amount of CT images in each validation dataset. A more detailed description of each dataset can be seen in the supplements.

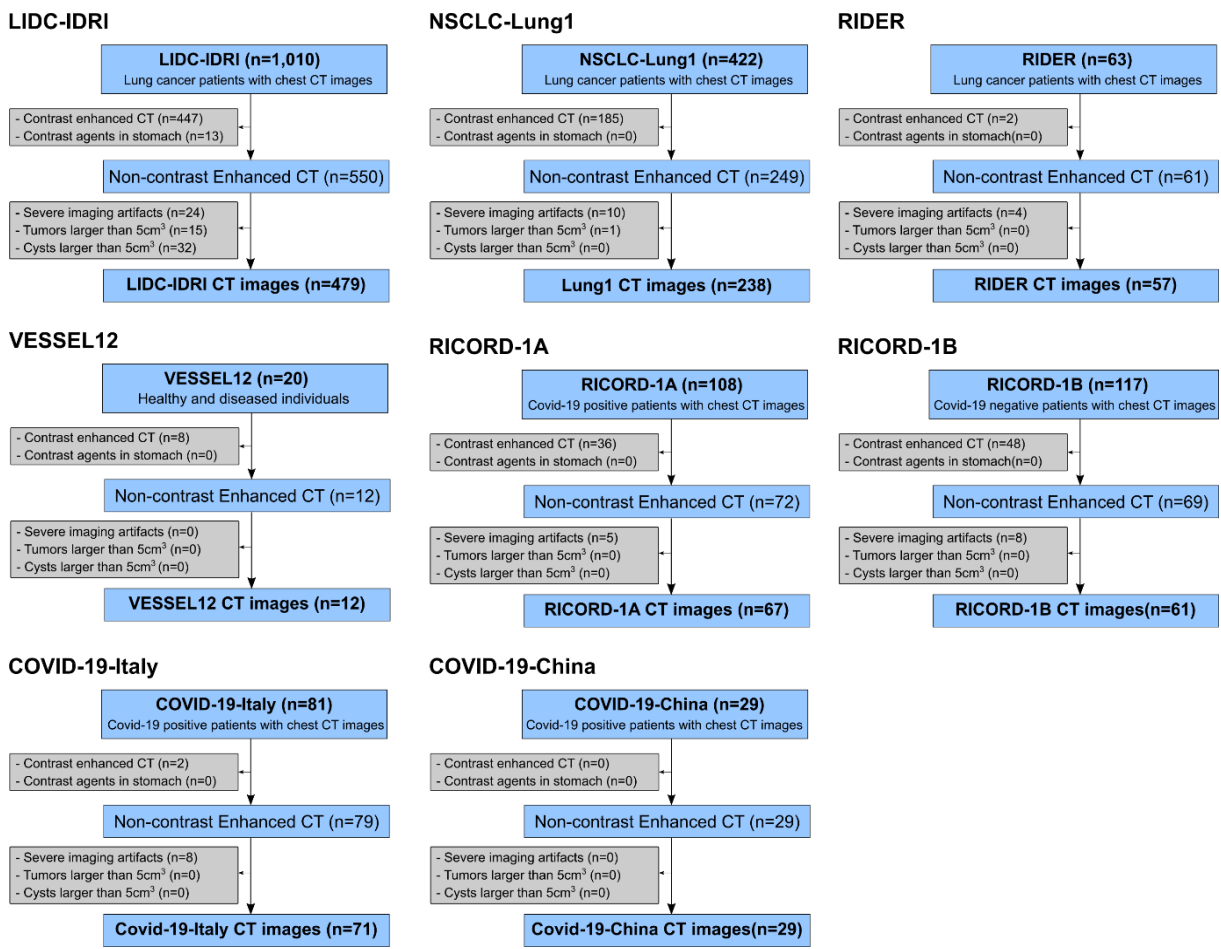

**Figure 2**: Consort diagram for each dataset of external validation.

**Data preprocessing**

CT images were reformatted to the Neuroimaging Informatics Technology Initiative (NIfTI) format. Linear interpolations were used to resample CT images to a common voxel spacing of 0.7 x 0.7 x 2.5 mm/pixel. CT Hounsfield units were windowed to [-200, 250] and rescaled to [0,1]. No data augmentation techniques were adopted during external validations.

**Expert Liver Segmentation and Liver Attenuation Measurement**

Imaged livers were segmented and attenuations were measured in HU on selected ROIs by experts. Manual segmentations were conducted independently by seven trained human experts and then validated/corrected by a radiologist (Y.H., 10 years of experience). Expert experiences varied from 1 data scientist and 6 medical doctors (1, 10, 10, 10, 10, and 15 years of experience). Liver contour lines were drawn on axial plain CT images of 2.5-mm thickness, from which the volumetric liver segmentations were constructed by the 3D Slicer software (Version 5). Liver fat content was estimated by liver attenuation and was measured on three separate 2D slices by selecting three circular ROIs with an area of around 2 cm$^2$ (Fig. S1). ROIs were placed on parenchymal regions to avoid heterogeneity such as hepatic veins, bile ducts, and focal nodules, on which the attenuations were measured to capture the full spectrum of liver fat. The attenuation was measured by experts without knowing the automated steatosis category. Hepatic steatosis was annotated as positive by experts if the liver attenuation <= 40 HU. Manual measurements of liver attenuations are demonstrated by a histogram in Fig. S2.

Liver attenuations were measured by only one expert on all CT images, except that 186 images were repeatedly measured to test agreements between four experts and automatic methods. Of these 186 images, a total of 86 images were randomly chosen from the LIDC-IDRI set, and the other 100 images were from COVID-19-China and COVID-19-Italy sets. Time was recorded to test the efficiency between experts and AI on the latter 100 images. The recorded time was spent loading images, sliding slices in the axial plane, selecting ROI on parenchymal regions, adjusting the area to around 2 cm$^2$, computing the mean voxel value, and saving ROI into NIfTI format. Expert experiences varied from 1 data scientist, 6 medical doctors (1, 10, 10, 10, 10, 15 years of experience) to 1 radiologist (Y.H., 10 years of experience).

**Deep Learning-enabled ROI Selections for Liver Attenuation Measurement**

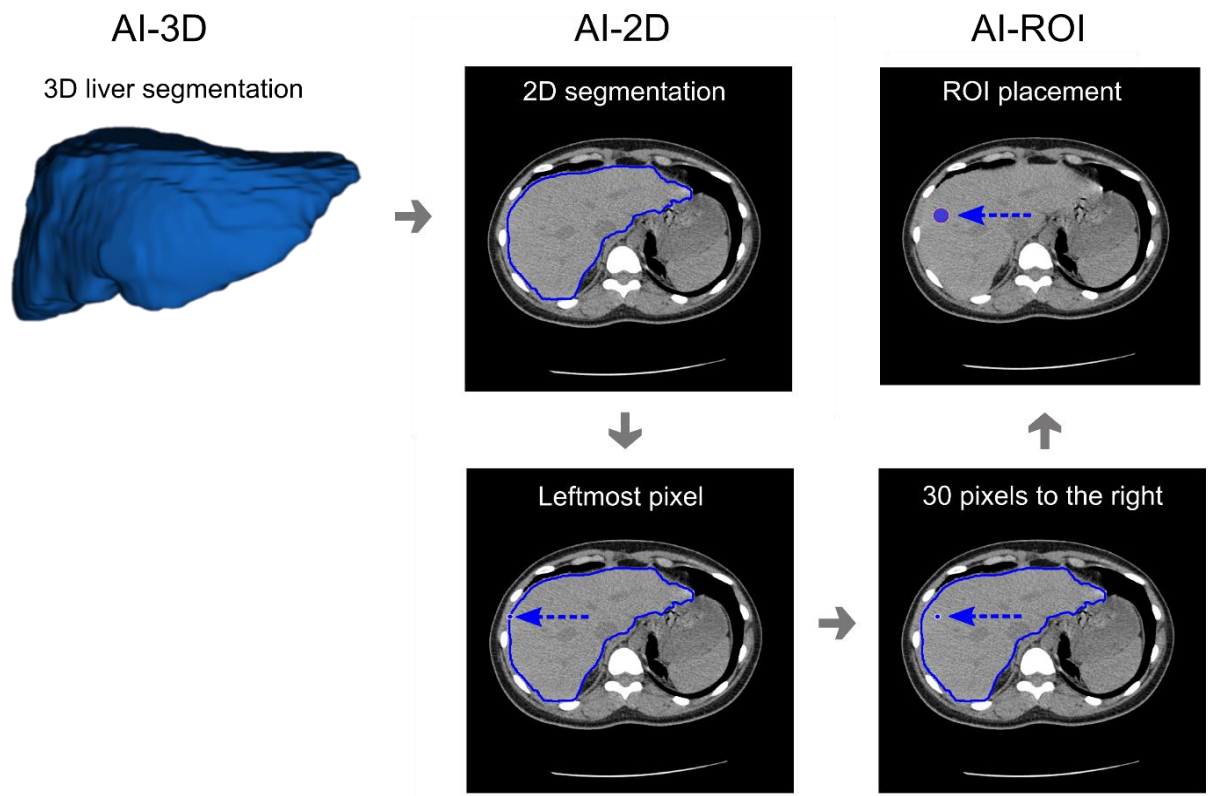

**Figure 3**: Liver attenuation was automatically measured by an existing deep learning system and two different post-pressing methods. **AI-3D:** AI-calculated mean voxel value directly by a deep learning system in a 3D manner; **AI-2D:** one post-processing method, AI-calculated mean pixel value on a single axial slice which has the biggest cross-sectional segmentation area out of the 3D segmentation; **AI-ROI:** another post-processing method, AI-calculated the mean pixel value on ROI which were automatically selected at three cross-sectional slices at different hepatic levels (blue). The center of circular ROI was set by 30-pixel-right to the leftmost voxel in each slice. Three slices were chosen as the AI-2D slice and two neighborhood slices with a distance of 5 mm. One center slice was shown here as an example. ROI = region of interest.

In AI-3D measurement, the liver attenuation was directly measured by the segmentation of a 3D U-Net[20] model without prior retraining (Fig.3). 3D U-Net was a deep learning model that was initially proposed for biomedical image segmentation[29]. The 3D U-Net model adopted in our study was developed for liver segmentations on asymptomatic heavy smokers screening for lung cancer[20]. The 3D U-Net model architecture and implementation details and are shown in the supplementary methods. In the model output, the largest connected component was retained by the connected-component analysis and designated as the segmented liver.

All voxels within the AI-segmented liver were computed to get a mean value as the AI-3D attenuation. Segmentation time, including model inferring and saving output array into NIfTI format, was recorded to compare the efficiency with experts.

AI-2D and AI-ROI are independent post-processing methods of the liver segmentation to derive the liver attenuation. AI-2D derives attenuation from the 2D pixel value of one single slice that has the biggest cross-sectional segmentation area. All pixels recognized as liver in this axial slice were computed to get an AI-2D attenuation. As another post-processing method, AI-ROI derives attenuation by calculating the mean pixel value on three ROIs placed by AI. Three axial slices were automatically chosen as one slice with the biggest cross-sectional segmentation area plus two neighborhood slices with a distance of 5 mm. The center of circular ROI was set by 30-pixel-right to the leftmost liver voxel in each slice. This optimized 30-pixel distance was set by experience to ensure the ROI was placed within the liver contour and also away from non-parenchymal regions. The area of ROI was set around 2.0 cm$^2$ to mimic the manual selection.

**Statistical Analysis**

AI-generated segmentations were compared to expert segmentations using the Dice coefficient, Jaccard coefficient, Hausdorff distance, and average symmetric surface distance with standard deviations (Python package: medpy.metric.binary, Version 0.4.0). In liver attenuation measurements, two-sided Kolmogorov-Smirnov tests (Python package scipy.stats, Version 1.5.3) were conducted to test if attenuations follow a normal distribution and if AI and expert measurements follow the same distribution. Spearman's correlation coefficient (Python package scipy.stats, Version 1.5.3) and intraclass coefficient (Statistical software SPSS version 25, two-way random absolute agreement, single measures) were calculated for the correlation between manual and automatic measurements. In steatosis categorization, the area under the curve (AUC), sensitivity, and specificity from receiver operating characteristics (ROC) were rated for the AI performance (Python package: sklearn.metrics, Version 0.23.2) and bootstrap sampling (n=1000) was used to derive the 95% confidence interval (CI) of categorization performances.

**Results**

**Dataset Characteristics**

Eight multicenter international datasets were used to evaluate deep learning-enabled detection for hepatic steatosis in three steps: (i) 3D liver segmentation; (ii) liver attenuation measurements; (iii) hepatic steatosis classification. In a total of 1,014 chest CT images, 786

(77.5%) CTs were scanned for lung cancer and 228 (22.5%) for COVID-19 prospectively. 931 (91.8%) were normal liver and 83 (8.2%) were hepatic steatosis annotated by experts. Table S1 summarizes the characteristics of each dataset.

**AI Performance on 3D Liver Segmentation**

The automatic segmentations were compared to expert segmentations and achieved a DSC of 0.957 ± 0.046 (mean ± std). 81.56% is greater than 0.95 and only 6.8% are below 0.90 in DSC results. DSC distributions are also grouped by normal and steatosis condition in box plots (Fig. 4). There is a clear trend toward better segmentation accuracy on normal livers: DSC is 0.959 ± 0.040, compared to 0.925 ± 0.081 on the liver with steatosis. Separate performances of each validation set in DSC, Jaccard Coefficient, Hausdorff distance, and average symmetric surface distance are shown in Table S1. More segmentation examples and outlier analysis can be seen in supplementary methods and figures.

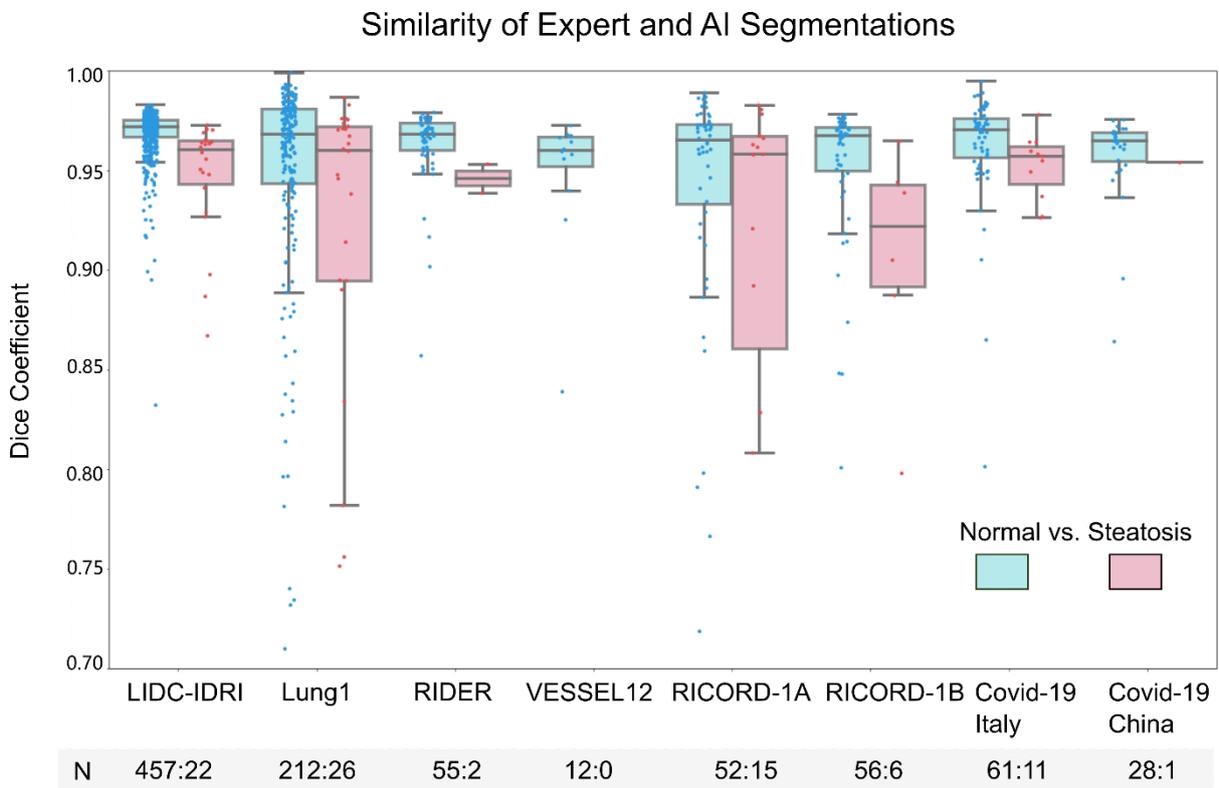

**Figure 4. A:** DSC between the deep learning system (AI) and expert segmentations grouped by expert annotations of normal and steatosis. The amount of CT images is denoted at the bottom (normal vs. steatosis).

**AI Performance on Liver Attenuation Measurement and Steatosis Classification**

All measurements of AI and experts do not follow normal distributions. In the attenuation comparison, AI-ROI is not significantly different from the expert measurements (P = 0.545). It can accurately select ROI on parenchymal regions even if the deep learning segmentation fails (Fig.S7). Both AI-3D and AI-2D yielded significantly different results from the expert (P < 0.001). Attenuation agreements and examples are shown in Fig. 5. Scatter plots of attenuation comparison and confusion matrix can be seen in Fig. S3&S4. In the steatosis classification, AUC, sensitivity, and specificity were shown in Table 1, along with the 95% confidence interval. While AI-ROI demonstrated same-to-expert accuracy in liver fat quantification, it is the AI-3D that achieved the best classification performance (p <0.001).

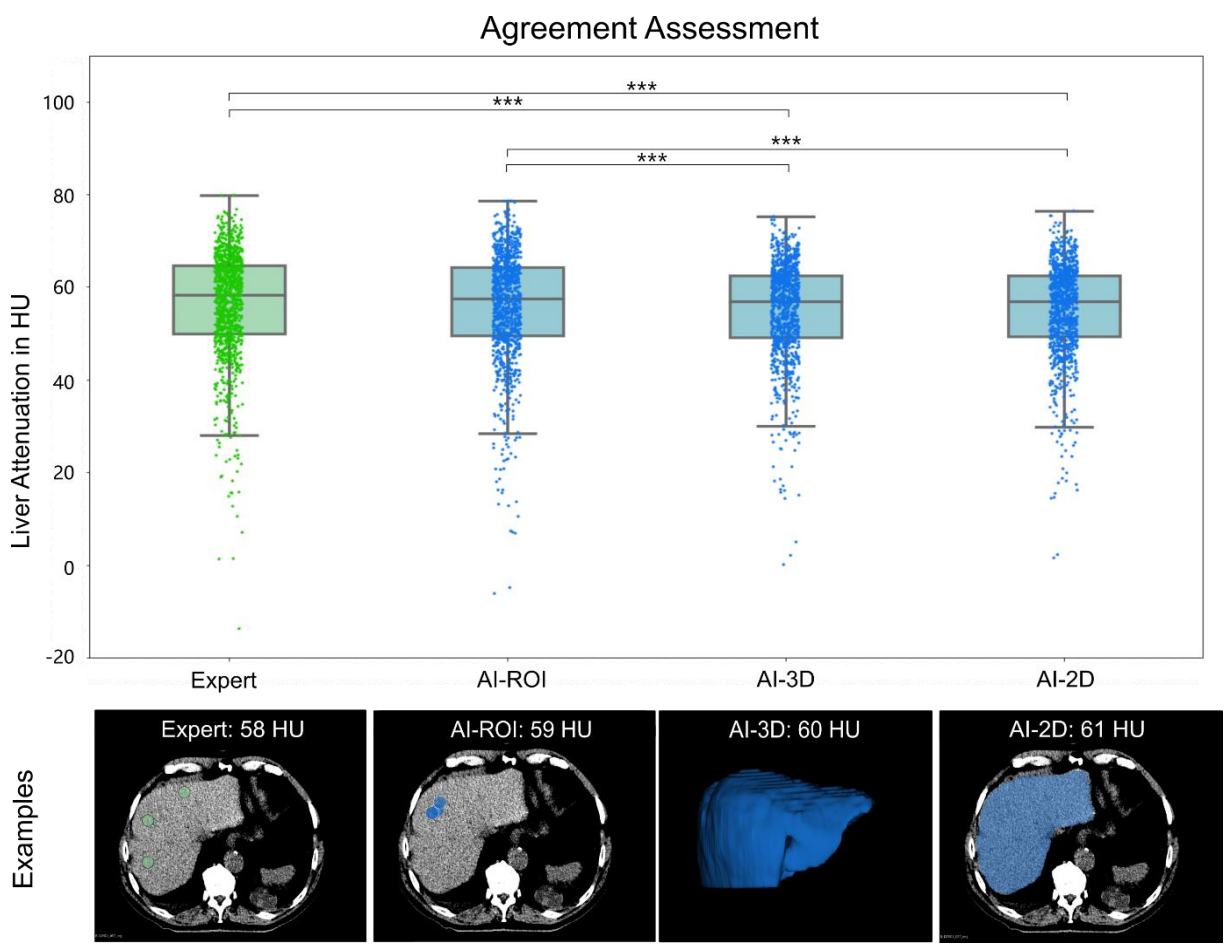

**Figure 5**: Box plots show liver attenuation measured by experts (green) and three automatic (blue) methods. Correlations are not labeled if differences are not significant. ***, p < 0.001; **, p < 0.01; *, p < 0.05. Examples of measurement methods are shown at the bottom, on which the attenuation is denoted. In expert and AI-ROI method, three ROIs should be placed in three different hepatic levels but are shown in one image here for a vivid visualization.

| Methods | AUC (95% CI) | Sensitivity (95% CI) | Specificity (95% CI) |
|---|---|---|---|
| AI-ROI | 0.921 (0.883 - 0.959) | 0.855 (0.779 - 0.929) | 0.987 (0.980 - 0.994) |
| AI-3D | 0.939 (0.903 - 0.973) | 0.880 (0.807 - 0.948) | 0.999 (0.997 - 1.000) |
| AI-2D | 0.894 (0.850 - 0.938) | 0.795 (0.709 - 0.882) | 0.994 (0.988 - 0.998) |

**Table 1:** Performances of AI methods in hepatic steatosis classification. 95% CI = 95% confidence interval; AUC = the area under curve.

**Liver Attenuation Concordance among Experts vs. AI Measurements**

The live attenuation was independently measured by four human experts and three automatic artificial intelligence (AI) methods, for which the consumed time was compared as well. Attenuation measurements of AI and human experts do not follow the normal distribution. AI-ROI measurements are not significantly different from all four experts. AI-3D and AI-2D are significantly different for the three experts (Z.Z., F.X., and Z.W.) but not for the other expert (N.Z.). The statistical differences between experts and AI methods can be seen in Fig. 6. A. Comparison of ROI placements between AI-ROI and experts are shown in Fig. S1, and AI-ROI placements along are shown in Fig. S7. In time assessment, significant differences exist between each expert and AI method, except AI-ROI vs. AI-2D (Fig. 6. B). Overall, AI methods were significantly faster than expert measurements (mean time 17.7 vs. 65.6 seconds). The separate time of each AI method and expert is shown in Fig. 6. B. Measurement time is even significantly different between different human experts.

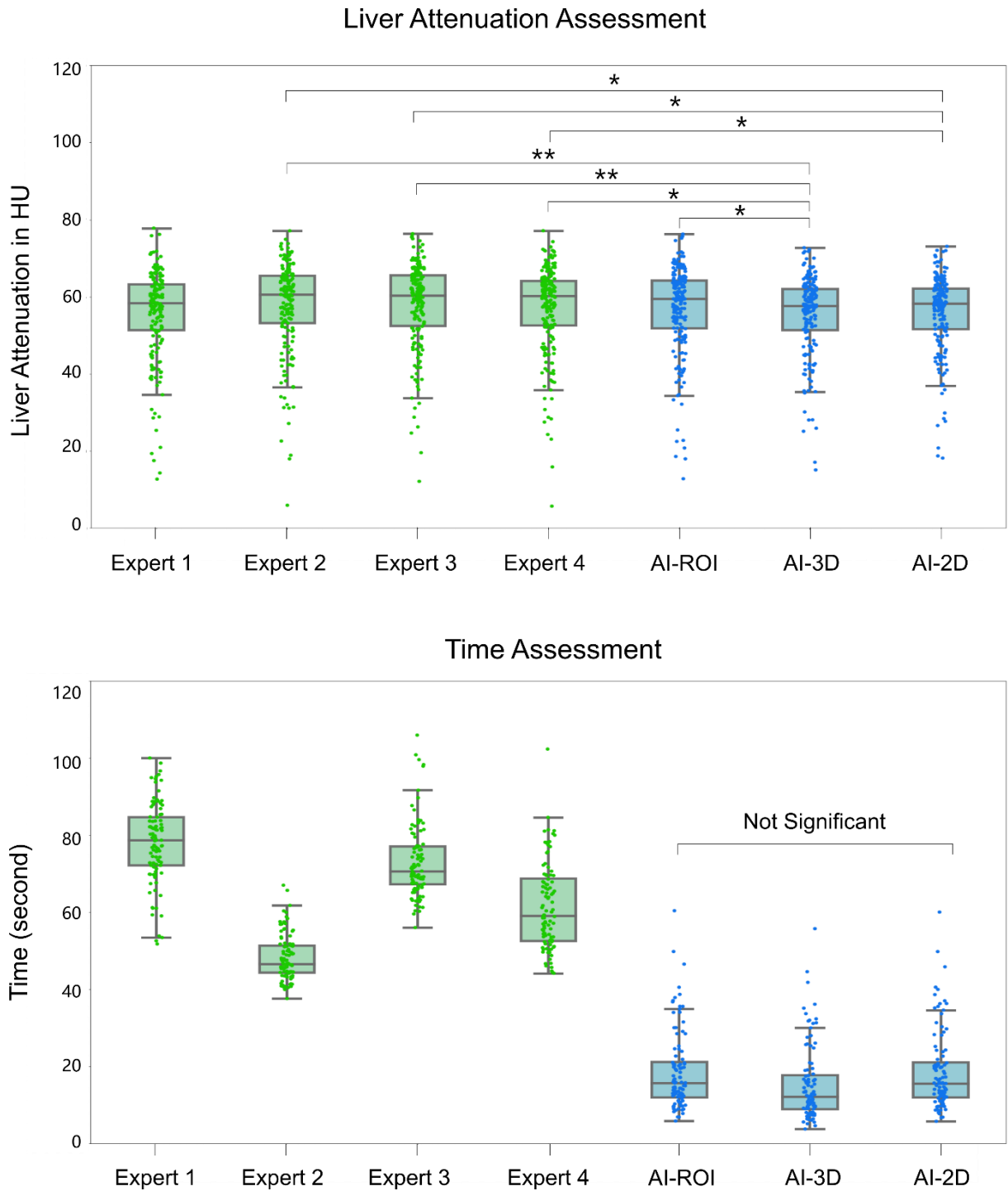

**Figure 6. A:** Liver attenuations were measured by four human experts and three automatic AI methods (blue) in 186 CT images. Four experts are N. Z., Z. W., Z. Z., and F. X. Signification differences were denoted with p value: ***, $p < 0.001$; **, $p < 0.01$; *, $p < 0.05$.
**B.** Time consumptions were recorded in a subset of 100 CT images. CT was measured by human experts (green) and AI methods (blue). Differences are significant ($p < 0.001$) between each expert and AI method, except AI-ROI vs. AI-2D ($p = 0.999$).

**Discussion**

In this study, we externally validated a deep learning system and used it to purpose a new post-processing method to automatically select ROI for hepatic steatosis detection on chest CT images. Our novelty is that the method achieved same-to-expert accuracy and reduced 70% time compared to expert measurements on CT images. We also release 1,014 labeled 3D liver segmentations of plain chest CT scans to promote deep learning segmentation algorithms. Our fully automatic method demonstrated robustness in multicenter multiparameter CT images that were scanned to determine other medical conditions, including lung cancer and COVID-19 diseases. The robustness is thus unbiased since CT was scanned without concerns of fatty liver beforehand. The method was successfully applied to 1,014 CT images, of which 8.2% were diagnosed with hepatic steatosis by experts, and it might represent a total of 50+ million individuals if scaled to international lung cancer and COVID-19 populations. Therefore, our method empowered by international validations holds the potential to identify millions of individuals at risk of hepatic steatosis and even enhance the current non-invasive diagnostics.

Our novel method, AI-ROI, is the first fully automatic and end-to-end method to imitate human experts selecting ROI to measure liver attenuation on CT images. Several automatic methods[17–20] reported good agreements with expert measurements, but significant differences still existed since the liver attenuation was computed in a 3D manner and non-parenchymal regions were not avoided. In comparison, our AI-ROI method best mimics the expert measurements by selecting ROI on the liver margin to avoid non-parenchymal regions. Though it is a post-processing method based on deep learning segmentation, it can automatically select ROI on liver parenchyma even if the segmentation is not successful. As a result, AI-ROI measurements were not significantly different from the expert measurements. The AUC of AI-ROI for hepatic steatosis classification is 0.921 (1 means perfect). We attribute the slight imperfection to the inter-reader variability within the ROI methodology since this near-perfect AUC of 0.921 was only compared to the CT reads from one single expert. Thus, the AUC of AI-ROI may be higher or lower, depending heavily on the opinions of an individual. To address this concern, we further compared AI-ROI to four human experts and demonstrated its persistent equivalences. Conclusively, we believe AI-ROI performance is comparable to human expert performance in hepatic steatosis classification. This is of great importance, as the new method could help to reduce labor-intensive manual work and inter-reader variability.

Liver fat was also automatically measured by an existing deep learning system (AI-3D) and another post-processing method (AI-2D). However, the liver attenuation measurements are statistically significantly different from that of experts'. The reason might be that AI-3D and AI-2D segmentations include non-parenchymal regions, while the ROI method samples a parenchymal portion. However, AI-3D achieves a statistically higher AUC, sensitivity, and specificity than AI-ROI in hepatic steatosis classification. The reason might be that AI-3D derives the attenuation in a 3D manner therefore not subject to sampling errors. Future studies are recommended to prove the correlation between volumetric attenuation and clinical diagnostics.

The current non-invasive, standard clinical imaging methods to measure liver attenuation includes ultrasound, MRI, and CT images[30]. In this study, we focus on non-enhanced chest CT images for two reasons: CT's high statistical correlation with hepatic steatosis[6,7] and the abundance of CT images at lung cancer[10,11,31] and COVID-19[12,32] datasets. While CT scans are continually ordered, CT's utility in detecting hepatic steatosis remains unexplored worldwide. In addition, we chose to validate this deep learning system[20] and use it to explore new methods because it first applied deep learning to liver segmentations on non-contrast enhanced chest CT scans. Other automatic methods either did not use deep learning[17,18] or did not use plain chest CT scans[19,33–35]. Overall, Deep-learning segmentations demonstrated high accuracies with our efforts to include large-scale external validation datasets, despite the slight decrease in accuracy from the internal[20] to our external validations (mean DSC 0.970 vs 0.957). The decrease might be attributed to different CT imaging parameters. We also observed that the segmentation accuracy in the steatosis group is mostly lower than in the normal liver. One reason might be that the deep learning model was trained mostly on normal livers and therefore biased. The other reason might be the different textures and brightness in CT images. Retraining the model with more livers with steatosis might correct the biased segmentation.

Our method contains four innovations that are immensely beneficial to clinical practice and AI developments: (i), our method is explainable: ROI placements and liver attenuation measurements are clearly visualized on CT images. Thus, the subsequent classification of hepatic steatosis is explainable to people who have no idea about deep learning; (ii), the method was proved to be accurate on eight distinct datasets, demonstrating generalizability across CT imaging protocols of different clinical centers as well as potential patient demographical differences, accounting for any potential regional, national, and racial differences. This one-size-fits-all generalizability would effectively minimize the concern of AI biases for clinical application. (iii), this method can boost clinical efficiency, as the AI requires mere seconds to measure liver attenuation, allowing clinicians to be able to review evidence of steatosis in 18.8 seconds. (iv), the expert segmentations contained herein are to be released to the public freely to promote deep-learning segmentation algorithms. Deep learning requires large datasets to fulfill its potential. Publicly available datasets have already promoted deep-learning algorithms on contrast-enhanced CT images by releasing only 200 liver segmentations[33] or 300 kidney segmentations[36]. As the largest open data, we believe that our released 1,014 liver segmentations would advance greatly the deep learning algorithm on non-enhanced chest CT images.

Despite the innovative aspects, limitations of this study still exist. For instance, hepatic steatosis could be also defined by a liver-to-spleen attenuation ratio of less than 1, in addition to the metric used in this study (liver attenuation of less than 40 HU); however, only the latter definition was used in this AI system. Further optimized AI systems may need to include the liver-to-spleen ratio. Another source of bias comes from the underlying patient population who, despite not necessarily being diagnosed with hepatic steatosis at presentation, are still mostly unhealthy due to lung cancer or COVID-19. To ensure the generalizability of this screening method, more validation datasets containing mostly healthy volunteers are necessary. Furthermore, the manual segmentations were conducted by 6 human experts but certified by only one radiologist. Having a cohort of experts in radiology may provide more convincing ground truths. Some other limitations of this AI system include its lack of prognostic value, as participant demographics and their underlying disease progression were not available.

In conclusion, we broadly validated a deep learning system for liver segmentation and used it to purpose a fully automated method to select ROI for hepatic steatosis detection on CT images. The novel method might allow fast and generalizable detection of hepatic steatosis across CT imaging protocols and clinical centers, as well as participants' regions, nations, and races. Plus explainability, it could be applied to clinical practice to detect and therefore prevent early-stage hepatic steatosis from progressing to end-stage liver failures.

## Data Availability

Our study materials are collected and reformatted from publicly available datasets under their copyright licenses (Table S1). The reformatted CT images and 3D expert segmentations can be freely downloaded by Google Drive https://drive.google.com/drive/folders/1-g_zJeAaZXYXGqL1OeF6pUjr6KB0igJX or Baidu Wangpan https://pan.baidu.com/s/1nRv-FJU4HtQ4nXi9H9145Q?pwd=2022 (passcode: 2022).

## Code availability

The deep-learning method is adopted from this study[20] and therefore not available here.

## Acknowledgments

The authors thank the support of Dr. Hugo Aerts for accessing the automatic method. This study was supported by a Fostering Foundation of the Second Hospital of Shandong University (Grant No. 2022YP46) and a Hainan Provincial Natural Science Foundation of China (Grant No. 821QN406).

## Author Contributions

**Contributions:** X.L. and Y.H. equally contributed to this project. Detailed author contributions are as follows: Figures: Z.Z.; Code implementation and execution: Z.Z.; Code reviewing: Z.Y.; CT segmentation and annotation: All authors; Study design: Z.Z., Y.H., X.L.; Test data collection: Z.Z.; Data analysis and interpretation: Z.Z., Z.Y., J.L., Y.H.; Critical revision of the manuscript for important intellectual content: All authors; Statistical Analysis: Z.Z., Y.H, X.L.; Study supervision: Y.H., X.L.

## Ethics Declaration

**Competing interests:** The authors declare no competing interests.

**Supplementary Methods**

**Datasets**

**LIDC-IDRI:** The Image Database Resource Initiative (IDRI) was created to further advance the Lung Image Database Consortium (LIDC) in 2004[21]. The LIDC-IDRI contains a total of 1018 chest CT scans from 1010 patients, including both contrast-enhanced and non-enhanced CT scans. Images were collected from seven participating academic institutions and eight medical imaging companies in the USA. LIDC consists of diagnostic and lung cancer screening chest CT scans with annotated lung lesions. It is originally used to develop automated lung cancer detection and diagnosis. LIDC images were constructed from 4 scanner manufacturers and 17 different CT imaging models. The tube peak potential energies used for scan acquisition ranged from 120 to 140 kV. Tube current ranged from 40 to 627 mA. Slice thicknesses ranged from 0.6 to 4.0 mm. The reconstruction interval ranged from 0.45 to 5.0 mm. The in-plane pixel size ranged from 0.461 to 0.977 mm. Each CT scan was initially presented at a standard brightness/contrast setting without magnification. No participant demographics (age, gender, etc.) or clinical information is available for this dataset.

**NSCLC-Lung1:** NSCLC stands for non-small cell lung cancer. The Lung1 data set was released in 2014 which consisted of 422 NSCLC patients in the Netherlands[22]. 132 are women and 290 are men. The mean age was 67.5 years (range: 33–91 years). Patients were included if they have confirmed diagnoses of lung cancer or underwent treatment with curative intent. This dataset was initially proposed to assess the prognostic value of radiomic features for lung cancer. CT scans and clinical data were available in this study[22].

**RIDER:** The RIDER data set consists of 31 patients with two CT scans acquired approximately 15 min apart[23]. Patients with non-small cell lung cancer were recruited in 2007 at Memorial Sloan-Kettering Cancer Center, New York, USA. The mean age is 62.1 years (range, 29-82 years), 16 were men (mean age, 61.8 years; range, 29-79 years) and 16 were women (mean age, 62.4 years; range, 45-82 years). Parameters for the 16-detector row scanner were as follows: tube voltage, 120 kVp; tube current, 299-441 mA; detector configuration. Parameters of the 64-detector row scanner were as follows: tube voltage, 120 kV; tube current, 298-351 mA.

**VESSEL12:** This dataset is collected from the VESsel SEgmentation in the Lung (VESSEL12) challenge held in 2012[24], which is to compare automatic methods of lung vessel segmentations taken from both healthy and diseased populations. CT scans were collected from three hospitals in the Netherlands and Spain in a variety of clinically common scanners and protocols[24]. The dataset released 20 CT scans and around 10 scans contain abnormalities such as emphysema, nodules, or pulmonary embolisms.

**MIDRC-RICORD-1A & 1B:** Medical Imaging Data Resource Center (MIDRC); RSNA International COVID-19 Open Radiology Database (RICORD); Release-1A: Chest CT COVID Positive (MIDRC-RICORD-1a); Release-1B: Chest CT COVID Positive (MIDRC-RICORD-1b)[25]. The data was collected in April 2020. Each dataset consists of 120 chest CT scans from four international sites: the USA, Turkey, Canada, and Brazil. The dataset has two inclusion criteria: 1. Adults underwent chest CT scans for suspected COVID-19 infection; 2. COVID-19 positive (1A) confirmed by one or more conditions: reverse-transcription polymerase chain reaction test, immunoglobulin M antibody test, or clinical diagnosis using hospital-specific criteria.

**COVID-19-Italy:** The dataset is originally made of 62 COVID-19-positive patients and then enriched to 81 patients[26,37]. The group of 62 patients underwent non-contrast chest CT scans in Italy in 2020. The average age was 56 years (range 20–83), and the male/female ratio was 23/27. Images were obtained with two different scanners with reconstructions of the volume at 0.3 to 1 mm slice thickness. Automatic lung tissue classification, clinical score, and intensive care unit information are provided as well. We chose the enriched set with one CT scan per patient, therefore adding up to 81 CT images in our study.

**COVID-19-China:** The dataset is made of 29 COVID-19-positive Chinese patients who received multiple non-contrast chest CTs between January 21st and April 12th, 2020 in Hubei Province, China[27]. The patients were predominantly female (69%, 20/29), and were 41 ± 10 years old (range 25 to 60 years old). Each patient underwent multiple CT scans at different time points. We chose the baseline CT scans per patient and therefore added up to 29 CT images in our study.

**Deep Learning Model Architecture and Implementation**

Our new post-processing method was based on a 3D U-Net model – added with residual connections on the original implementation[20,38] – for liver segmentations and liver attenuation measurements in a 3D manner. The 3D U-Net model was developed and evaluated on 551 non-enhanced chest CT images, which were randomly selected from the National Lung Screening Trial (NLST)[39]. As a randomized controlled trial for lung cancer screening, the inclusion criterion of NLST is asymptomatic heavy smokers.

Model architecture was shown in Fig. S5. The input dimension was $64^3$ and the output dimension was $2 * 64^3$. Convolutional layers with the stride of 2 were utilized to increase and decrease each stage's dimensions. The kernel's size was 3*3*3 in all convolutional layers. In addition, parametric rectified linear units and batch normalization were adopted in each down-sampling and up-sampling step. Dice loss function was adopted for the backpropagation algorithm. No prior retraining or data augmentation were adopted in implementations on our study. The 3D U-Net model was implemented in a Linux workstation with a P100 GPU using Pytorch (version 1.6.0)[40] and MONAI framework (version 0.4.0)[41].

**Outlier Analysis**

As the result of deep learning segmentation, DSC distributions are shown in a histogram (Fig. S6), among which 81.56% is greater than 0.95 and only 6.8% are below 0.90. Incorrectly segmented livers are shown in Fig. S7. Segmentation failures might attribute to CT imaging parameters at different datasets across the world. Even when the segmentation DSC is below 0.90, the liver contours are still recognized. Since the automatic ROIs are placed around the leftmost voxel of deep learning segmentations, AI-ROI placements are still correct and away from non-parenchymal regions.

**Supplementary Tables**

| Dataset | Initial Purpose | License | Location |
| --- | --- | --- | --- |
| LIDC-IDRI | Lung cancer | CC BY 3.0 | USA |
| NSCLC-Lung1 | Lung cancer | CC BY 3.0 | Netherlands |
| RIDER | Lung cancer | CC BY 3.0 | USA |
| VESSEL12 | Lung vessel segmentation | Not specified | Netherlands, Spain |
| RICORD-1A | Covid positive | CC BY-NC 4.0 | USA, Canada, Turkey, Brazil |
| RICORD-1B | Covid negative | CC BY-NC 4.0 | USA, Canada, Turkey, Brazil |
| Covid-19-Italy | Covid positive | CC BY-NC 4.0 | Italy |
| Covid-19-china | Covid positive | CC BY 4.0 | China |

**Table S1**: The amounts of images of interest were denoted for each dataset. Four datasets were initially obtained to diagnose COVID-19, three datasets were scanned for lung cancer detection, and one dataset was released for lung vessel segmentation. These CT images were scanned from eight countries and released by copyright licenses. CC-BY refers to the Creative Commons Attribution license and allows readers to distribute, remix, adapt, and build upon the material in any medium or format, so long as attribution is given to the creator. NC means for noncommercial purposes only.

| Dataset | Dice Coefficient | Jaccard Coefficient | Hausdorff Distance (mm) | ASSD (mm) |
|---|---|---|---|---|
| LIDC-IDRI | 0.967 ± 0.015 | 0.937 ± 0.027 | 21.391 ± 27.202 | 0.932 ± 1.434 |
| Lung1 | 0.942 ± 0.070 | 0.898 ± 0.107 | 28.605 ± 43.447 | 1.611 ± 3.660 |
| RIDER | 0.962 ± 0.020 | 0.927 ± 0.036 | 19.817 ± 12.230 | 0.970 ± 0.793 |
| VESSEL12 | 0.948 ± 0.035 | 0.903 ± 0.059 | 59.155 ± 85.549 | 3.668 ± 5.868 |
| RICORD-1A | 0.930 ± 0.083 | 0.879 ± 0.123 | 26.472 ± 16.907 | 1.626 ± 1.883 |
| RICORD-1B | 0.948 ± 0.041 | 0.903 ± 0.068 | 21.144 ± 14.934 | 0.903 ± 0.068 |
| COVID-19-Italy | 0.957 ± 0.023 | 0.925 ± 0.048 | 27.754 ± 20.072 | 1.222 ± 0.980 |
| COVID-19-China | 0.960 ± 0.028 | 0.918 ± 0.040 | 18.668 ± 8.694 | 0.917 ± 0.420 |
| **Total** | **0.957 ± 0.046** | **0.920 ± 0.071** | **24.131 ± 31.253** | **1.210 ± 2.255** |

**Table S2:** AI Liver segmentation performances on each dataset separately and conclusively. ASSD = average symmetric surface distance.

| Measurement | Expert | AI-ROI | AI-3D | AI-2D |
|---|---|---|---|---|
| LIDC-IDRI | 60.80 (n = 479) | 60.62 (p = 0.69) | 59.20 (p < 0.001) | 59.32 (p < 0.001) |
| NSCLC-Lung1 | 48.96 (n = 238) | 48.75 (p = 0.96) | 48.29 (p = 0.22) | 48.80 (p = 0.22) |
| RIDER | 61.16 (n = 57) | 60.32 (p = 0.98) | 57.73 (p = 0.003) | 57.87 (p = 0.01) |
| VESSEL12 | 57.15 (n = 12) | 56.09 (p = 1.00) | 55.62 (p = 0.87) | 55.23 (p = 0.87) |
| RICORD-1A | 49.38 (n = 67) | 47.56 (p = 0.86) | 48.81 (p = 0.73) | 48.73 (p = 0.73) |
| RICORD-1B | 55.85 (n = 61) | 54.70 (p = 0.52) | 54.44 (p = 0.39) | 54.32 (p = 0.28) |
| COVID-19-Italy | 53.28 (n = 71) | 52.49 (p = 0.96) | 51.69 (p = 0.36) | 52.05 (p = 0.48) |
| COVID-19-China | 57.55 (n = 29) | 56.71 (p = 1.00) | 56.57 (p = 0.57) | 55.92 (p = 0.37) |
| **Total** | **56.33 (n = 1014)** | **55.86 (p = 0.55)** | **54.94 (p < 0.001)** | **55.11 (p < 0.001)** |

**Table S3:** The mean liver attenuation of the expert's and AI's measurements on each dataset (n denotes the amount of CT images). Each group of AI measurements was compared to the expert measurement. 'Total' means the mean attenuation on all datasets.

**Supplementary Figures**

# ROI Placements: Experts vs. AI-ROI

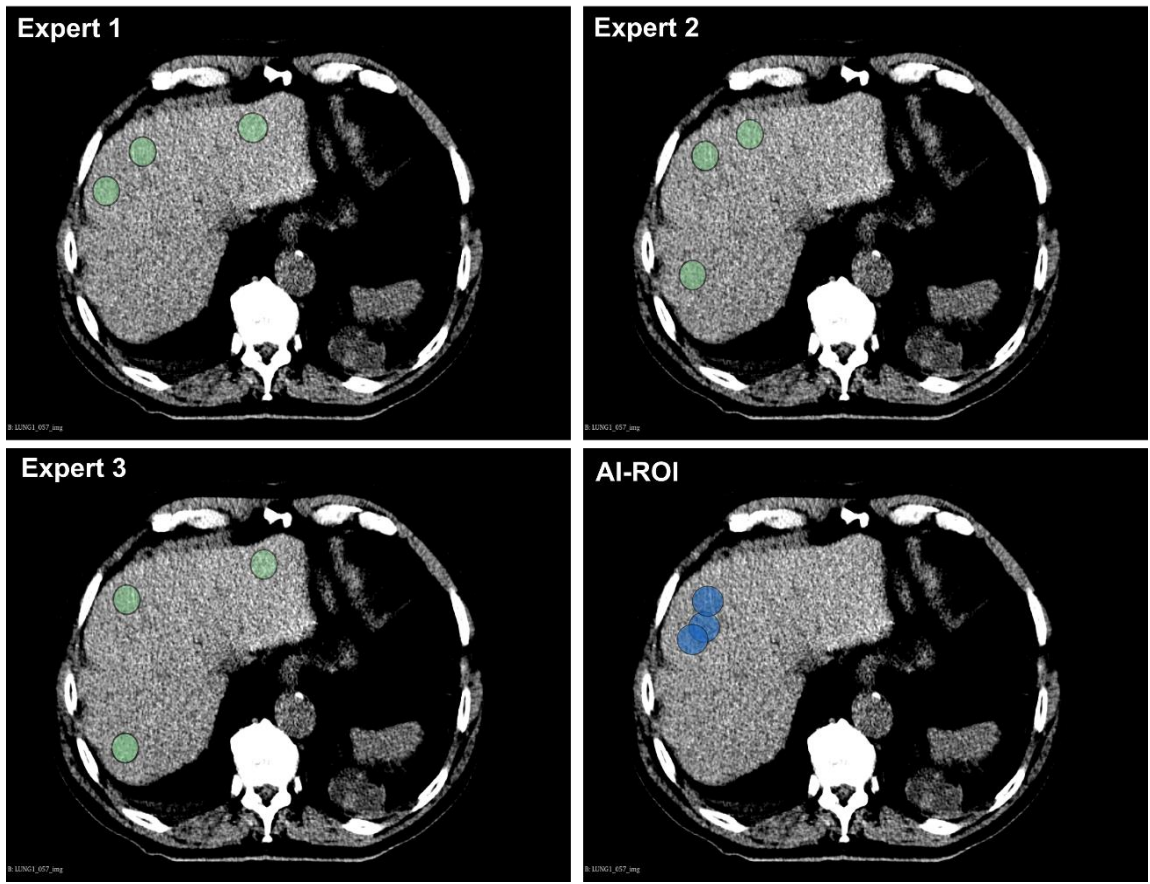

**Figure S1**: The attenuation of one example CT image was measured by three human experts (green) and AI-ROI (blue). Three regions were placed in three different hepatic levels but were shown in one image here for a vivid visualization.

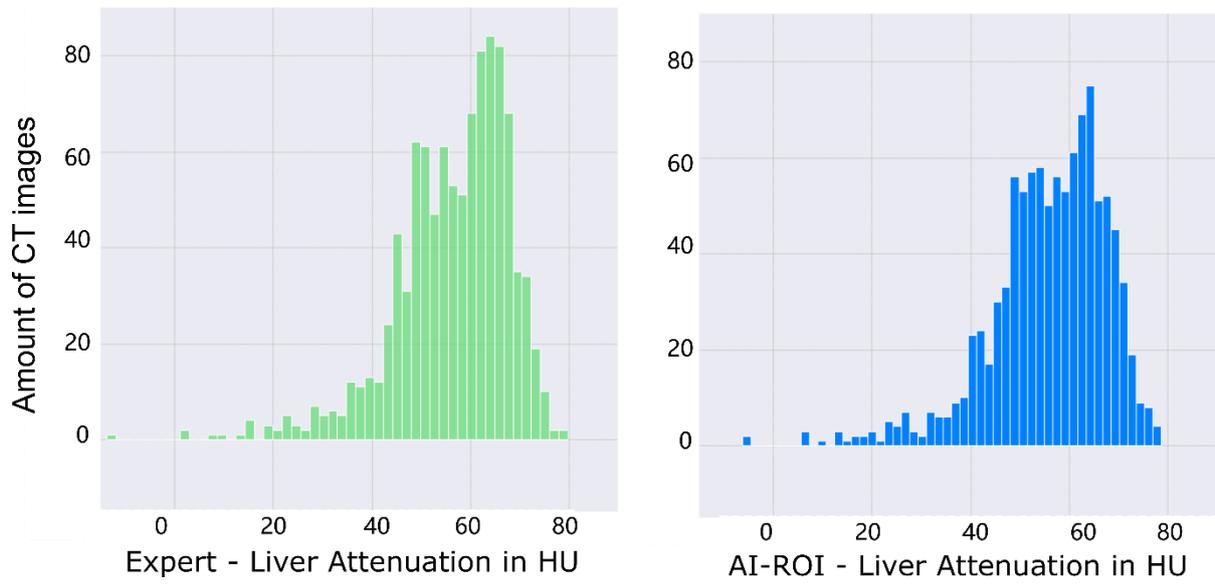

**Figure S2**: Histogram plots show distribution of liver attenuation measured by experts and AI-ROI method. HU = Hounsfield unit.

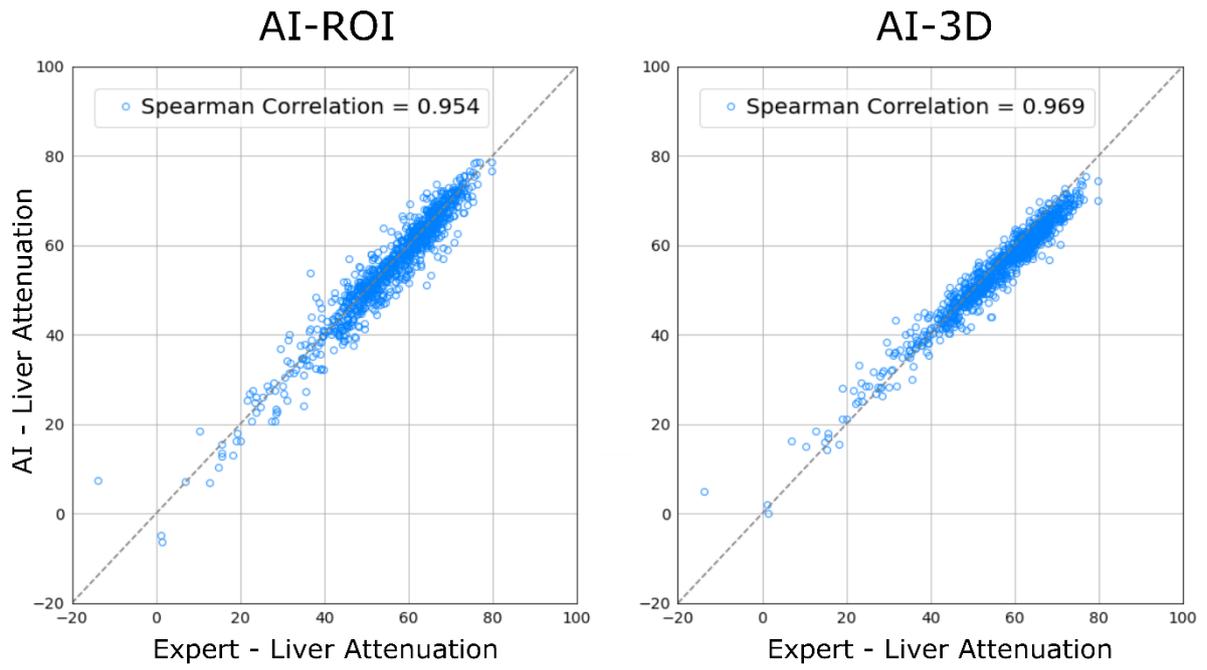

**Figure S3**: Scatter plots show liver attenuation measured in HU by AI methods and experts. Sperman correlation was denoted at the top.

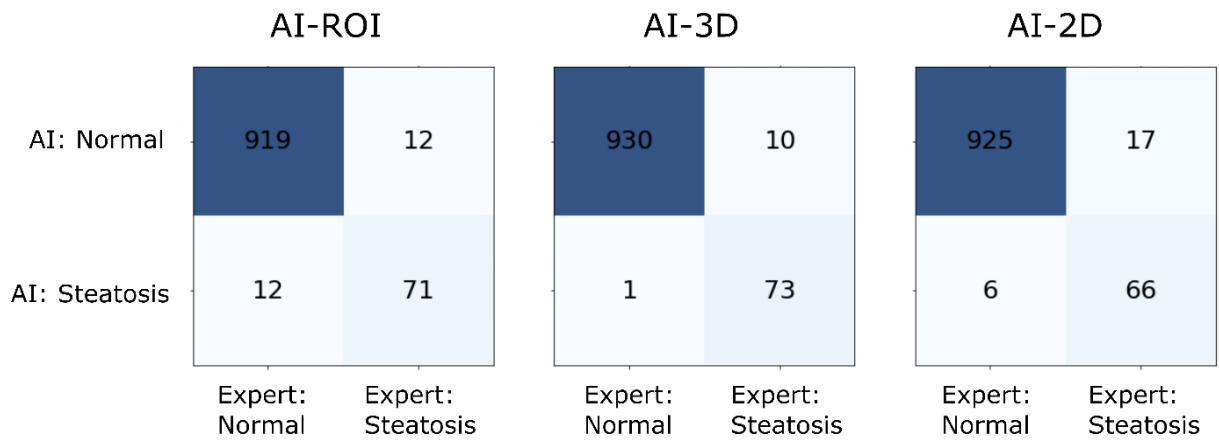

**Figure S4**: Confusion matrices were conducted to compare the steatosis categorization by AI methods and expert readers.

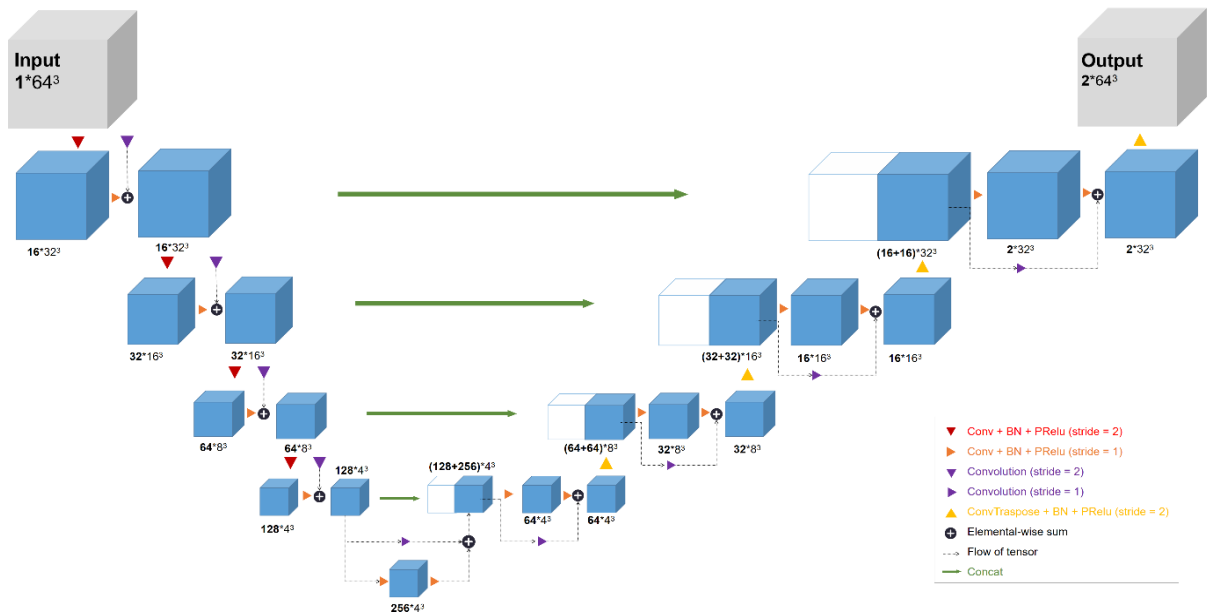

**Figure S5**: The model architecture of the 3D U-Net for volumetric liver segmentation. Prelu= parametric rectified linear unit, BN = batch normalization.

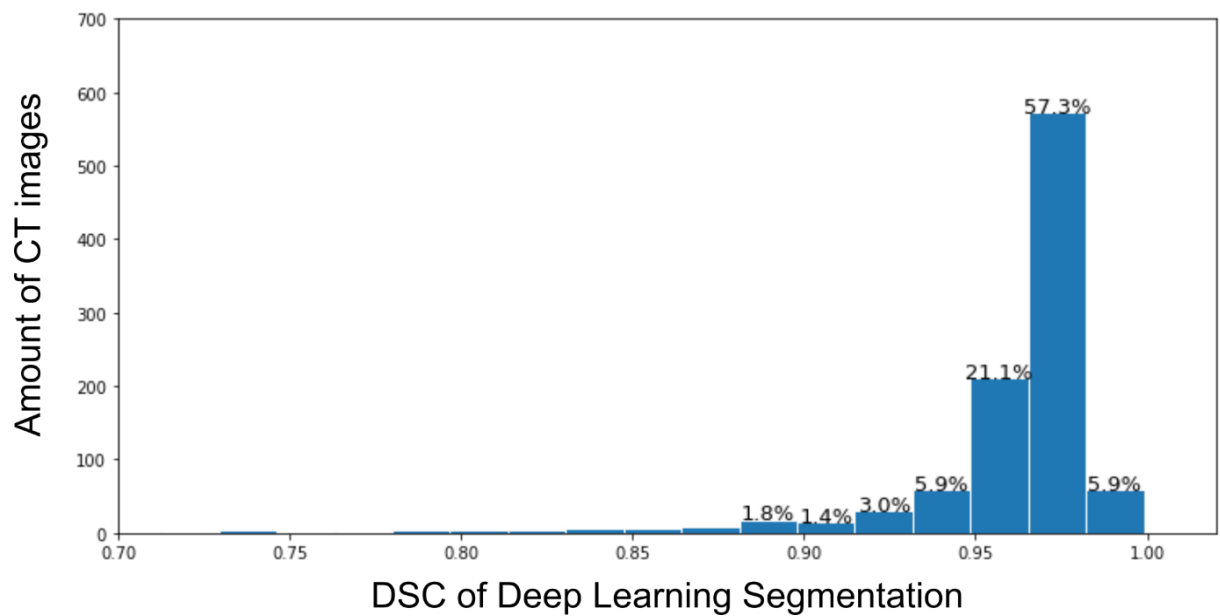

**Figure S6**: Histogram of DSC distributions on all CT images. DSC percentage is denoted at each bin if above 1%.

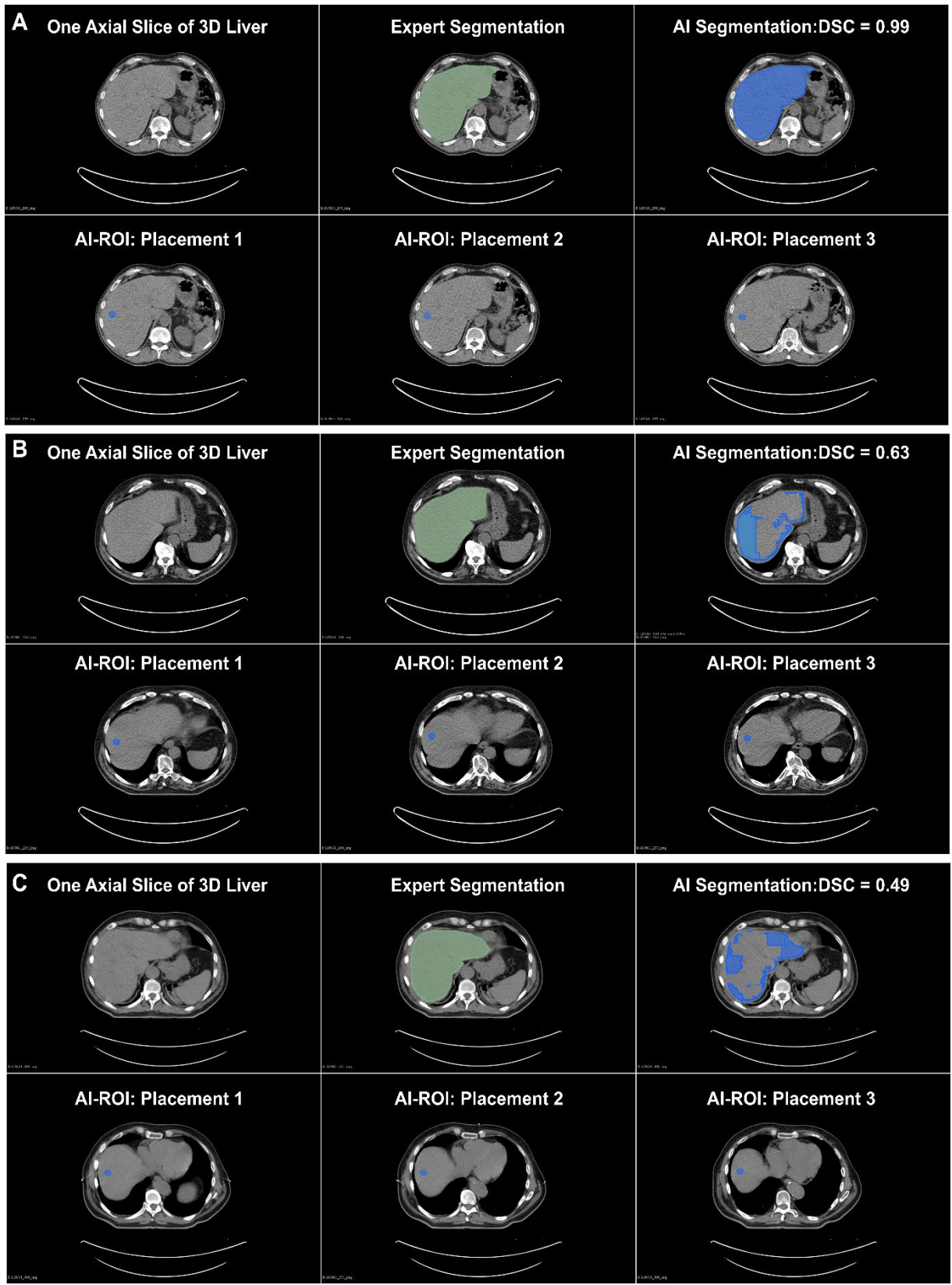

**Figure S7**: Segmentation and AI-ROI placement examples, including two outliers (B&C). Each example image was sliced from the CT scan with the AI (blue) and expert (green) segmentations. DSC = Dice coefficient.